\documentclass{elsart}

\usepackage{graphicx}

\usepackage{amssymb,amsmath,bbm}

\usepackage[latin1]{inputenc}

\begin{document}

\begin{frontmatter}

\title{Gauss map and Lyapunov exponents of interacting particles in 
        a billiard}

\author{Cesar~Manchein and}
\author{Marcus~W.~Beims\corauthref{cor}}
\corauth[cor]{Corresponding author}
\ead{mbeims@fisica.ufpr.br}

\address{Departamento de F\'{\i}sica,
             Universidade Federal do Paran\'a,\\
             81531-990 Curitiba, PR, Brazil}
\newcommand{\eq}[1]{Eq.~(\ref{#1})}

\begin{abstract}
We show that the Lyapunov exponent (LE) of periodic orbits with Lebesgue
measure zero from the Gauss map can be used to determine the main qualitative 
behavior of the LE of a Hamiltonian system. The Hamiltonian system is a 
one-dimensional box with two particles interacting via a Yukawa potential and 
does not possess Kolmogorov-Arnold-Moser (KAM) curves. In our case the
Gauss map is applied to the mass ratio ($\gamma=m_2/m_1$) between particles.
Besides the main qualitative behavior, some unexpected peaks in the $\gamma$
dependence of the mean LE and the appearance of `stickness' in 
phase space can also be understand via LE from the Gauss map.
This shows a nice example of the relation between the ``instability'' of the
continued fraction representation of a number with the stability of 
non-periodic curves (no KAM curves) from the physical model. Our results also 
confirm the intuition that pseudo-integrable systems with more complicated 
invariant 
surfaces of the flow (higher genus) should be more unstable under perturbation.
\end{abstract}


\end{frontmatter}

\section{Introduction}

The Gauss map \cite{Mane,Corless1,Corless2} is a chaotic map which generates 
the integers from a simple Continued Fraction (CF) representation of a real 
number. CFs have been used in several different scientific context like, for
example, the renormalization group theory \cite{Reno,Fuchss,Morrison}, 
expansion technique
applied to a model for Bloch electrons in a magnetic field \cite{Bloch},
stability of elementary particles \cite{Naschie} and their mass ratio 
representation \cite{Naschie,Marek}. In the context of nonlinear 
dynamics it has been applied to compute stable and unstable directions of 
maps \cite{SU}, to approximate irrational winding numbers for the 
KAM tori \cite{KAM} and to determine critical 
parameter values at which the KAM tori break \cite{Greene1,Greene2,satija}.

In this paper we discuss the stability of perturbed tori in a different 
situation. The unperturbed system is the problem of two particles in a 1D-Box 
interacting via Hard Point-like Collisions (HPC). Depending on the mass 
ratio between particles, the system can be integrable (Invariant Surface (IS) 
of the flow is a torus with genus $g=1$), pseudo-integrable (IS has a more 
complicated topology with $g>1$) and ergodic (see 
\cite{Richens,gorin,Gutkin}). Since the HPC case is linearly unstable, 
the LE is zero for each mass ratio and little is known about the stability 
of the IS. Since the Hamiltonian is not differentiable at the boundary
and at the collisions, no KAM curves exist for any perturbation.
Classical \cite{Richens} (quantum \cite{Richens,gorin}) 
results strongly suggest that the classical dynamics (level statistics) is 
more unstable (Gaussian Orthogonal Ensemble (GOE) distribution) for the 
pseudo-integrable than the integrable 
cases. Here the perturbed system is obtained by changing the HPC to a Yukawa 
interaction (YI). Recent classical results \cite{nos} also suggest that 
pseudo-integrable IS from the HPC are more unstable than the integrable ones
when the YI is turned on. We show here that the chaotic dynamics in the 
perturbed system, the 1D-Box with YI, is directly
related with the chaotic property of the {\it number} related to the mass ratio
$\gamma=m_2/m_1$ between particles. To do this we compare the LE from the 
infinite continued fraction representation of $\gamma$ with the maximal LE
from the two interacting particles in the 1D box. In fact, we show that the
qualitative behavior of the LE in the YI case can be reproduced qualitatively
using the LE from the Lebesgue Measure (LM) zero Periodic Orbits from Gauss Map
(POGM). We also show that the dynamics in the Yukawa case is more unstable 
when POGM are closer to non-POGM. In these cases `stickiness'
(tendency of orbits to get trapped) appears
more often in phase space. This is shown by using the most probable Lyapunov 
exponent, proposed recently \cite{nos} as a very sensitive tool to probe 
globally details in phase space dynamics.

It is well known from the KAM theorem that in two-dimensional systems the 
torus surface does not exist if $\alpha=\omega_1/\omega_2$ lies in a region 
for rational numbers, where $w_1,w_2$ are the frequencies of the unperturbed
problem. The region for rational numbers increases as the perturbation 
parameter increases. Therefore, only those irrationals that are hardest to 
approximate by rationals will yield the KAM surface. 
The residue criterion \cite{Greene2} establishes a correspondence between
the existence of a KAM curve and the stability of the periodic orbits that
approximate it. In other words, the stability of {\it periodic orbits} very 
close to {\it non-periodic orbits} (KAM curves), allows to make some 
statements about the destruction of the KAM curve. In our case the situation 
is different, the LE calculated for the {\it periodic orbits} with LM zero 
from the Gauss map, i.~.e.~the  {\it non-periodic curves} from the physical 
model (not KAM curves), allows us to get 
some insight about the stability of the IS from a linear unstable system
with LE zero. Therefore we have a {\it direct relation between the 
invariant surfaces with zero LEs and their stability under perturbation}.

In section \ref{1Dpoint} the main results from the 1D-Box HPC are summarized.
Section \ref{Sgauss} analyzes those properties of the Gauss map which are
relevant for the purpose of this paper. We calculate explicitally the LE for 
the POGM with LM zero which differ from the ergodic result. 
Section \ref{1Dyukawa} introduces
the smooth Yukawa interaction between particles and the LE distribution is 
calculated. The relation between the mean LE from the physical model and 
the LE from the POGM is demonstrated. The paper ends with
conclusions in section \ref{conclusion}.

\section{Two particles in a 1D-Box with hard point-like collisions}
\label{1Dpoint}

Two particles in a 1D-box with HPC can be treated as a particular case of
the motion of three particles on a finite ring \cite{triangle,triangle2},
which can be mapped onto the motion of a particle in a triangle billiard
\cite{2dtriangle}. The whole dynamics can be monitorated by changing the 
angles of the triangle billiard. These angles are functions of the masses 
ratio between particles.
It is possible to show \cite{casati} that the dynamics is non-ergodic if 
$\theta$ is a rational multiple of $\pi$, where

\begin{equation}
\cos{(\theta)}=\frac{1-m_2/m_1}{1+m_2/m_1}=\frac{1-\gamma}{1+\gamma}.
\label{cost}
\end{equation}
Writing $\theta=\frac{m}{n}\pi$, where $m$ and $n$ are integers, at most $4n$
distinct velocity values occur. These are the periodic orbits from the problem.
Although there are infinite mass ratios which give rational values of 
$\theta/\pi$, some of them are special: the integrable cases \cite{Koslov}
 $\gamma=1,3$ (or $1/3$), which have $\theta=\frac{1}{2}\pi$ and 
$\theta=\frac{2}{3}\pi$ (or $\pi/3$), respectively. These are the cases
for which the genus $g=1$ \cite{genus} (the IS of the billiard flow is a 
torus). For all other rational $\theta/\pi$ the dynamics is pseudo-integrable 
\cite{triangle2}, the invariant flow is not a torus  and  
gets more and more complicated as $g$ increases  ($1<g<\infty$) \cite{Gutkin}.
It was shown \cite{gorin} that for small $g$ the spectral statistics is close 
to semi-Poisson and it approaches the GOE statistics when $g$ is increased.
Table \ref{tab} shows some rational angles for the right triangle billiards 
with $g\le3$ 
and their relation with the mass ratio $\gamma$ and $\theta/\pi$ obtained from 
Eq.~(\ref{cost}). For genus $g=2$, for example, the values of $\theta$ are:
$\frac{1}{4}\pi,\frac{1}{5}\pi$ and the mass ratio are respectively
$\gamma\sim0.17$ and $\gamma\sim0.11$. 
On the other hand, when $\theta$ is an irrational multiple of $\pi$,
the velocities become uniformly dense \cite{arnold} in velocity space.
As a consequence, it is at least {\it possible} for the two-particle
with HPC  to be ergodic in velocity space.

\begin{table}[htb] 
  \centering
  \large
    \begin{tabular}{|c|c|c|c|c|c|c|}
       \hline
       genus~$(g)$ & $p/q$ & $\theta/\pi$ & $\gamma$ & $1/\gamma$&
         $\lambda^G(x_{k,800})$ & $x_{k,800}$(CF) \\
       \hline
     $1$ &  $1/4$ & $1/2$ & $1$ & $1$ & $7.37$ & $[\overline{2,800}]\sim1/2$ \\
       \cline{2-6}
      \mbox{Integrable}   & $1/6$ & $1/3$ & $1/3$ & $3$ & $7.78$ &
                                          $[\overline{3,800}]\sim1/3$    \\
       \hline
        $2$ & $1/8$  & $1/4$ & $0.17$ & $5.89$& $8.07$ & $[\overline{4,800}]
            \sim1/4$ \\
       \cline{2-6}
    \mbox{Pseudo-Int.}
  & $1/10$ & $1/5$ & $0.11$ & $9.47$ & $8.30$ & $[\overline{5,800}]\sim1/5$  \\
       \cline{2-6}
       \hline
   $3$ & $1/12$ & $1/6$ & $0.07$ & $13.9$ & $8.47$ & 
  $[\overline{6,800}]\sim1/6$ \\
       \cline{2-6}	
    \mbox{Pseudo-Int.}
 & $1/14$  & $1/7$ & $0.05$ & $19.2$ & $8.64$ &$[\overline{7,800}]\sim1/7$ \\
       \hline
    \end{tabular}
\caption{Some rational angles $p/q$ (see \cite{gorin}) for genus 
$g\le3$ from the triangle billiards and their relation 
with the mass ratio $\gamma$ and $\theta/\pi$ obtained from 
Eq.~(\ref{cost}). $\lambda^G(x_{k,800})$ is the LE obtained from period-2 
orbits of the Gauss map 
calculated very close (on the left) to $\theta/\pi$. The last row shows the
CF representation of the $x_{k,800}$ used to calculate the LE.} 
\label{tab}
\end{table}

Since the 1D-Box HPC is linear unstable \cite{zero}, all LEs
are zero and little is known about the stability of the above mentioned
ISs. In this work we  propose to analyse the stability of the 1D-Box HPC
using results from the Gauss map. We will show that, associating the 
irrational values of $\theta/\pi$ (and therefore $\gamma$) with properties 
of the LE for the LM zero POGM, we are able to make some statements about 
the stability of the ISs with LE {\it zero} from the HPC case.

\section{The Gauss map}
\label{Sgauss}

In this section the main properties of the Gauss map will be reviewed and
some numerical calculations for the LE will be performed. For more details 
we refer to the works of Corless et al \cite{Corless1,Corless2} and 
references therein. The Gauss map for $x$ in the interval $(0,1)$ is given by

\begin{equation}
G(x)=\frac{1}{x}-\left[\frac{1}{x}\right].
\label{gauss}
\end{equation}
The notation $[.]$ means to take the fractional part. The LE exponent
can be calculated from

\begin{equation}
\lambda^G(x)= \displaystyle\lim_{j\rightarrow\infty}\frac{1}{j}ln
      \left(\displaystyle\prod_{i=1}^j|G^{\prime}(x_i)|\right),
\label{LE}
\end{equation}
whenever this limit exists, where 
$G^{\prime}(x_i)=\partial G(x_i)/\partial x_i$. The Gauss map 
generates the numbers $n_1,n_2,n_3,\ldots$ from the simple CF
representation of a real number

\begin{equation}
x=n_0+\frac{1}{n_1+\frac{1}{n_2+\frac{1}{n_3+\ldots}}},
\label{CF}
\end{equation}
where the $n_i$ are all positive integers, except $n_0$ which may be zero or
negative. Here we consider $n_0=0$.  
The CF is represented in the form $x=n_0+[n_1,n_2,n_3,\ldots]$.
For rational values of $x$ the sequence of $n_i$ is finite and the LE from
Eq.~\ref{LE} cannot be calculated. For irrational
$x$ the sequence is infinite. For irrational quadratic values of $x$, the
sequence of $n_i$ are periodic, as exemplified below. The Gauss map is ergodic 
and for almost all initial conditions the LE may be calculated 
explicitly by $\lambda^G(x)=\pi^2/6\log{2}=2.3731\ldots$. {\it 
Periodic orbits} in the Gauss map occur when a sequence of integers 
repeat. A fixed point 
has the property $x_{n_1}=[n_1,n_1,n_1,\ldots]$, which is written in a more 
simplified manner as $[\overline n_1]$.  For 
period-1 periodic orbits the LE can also be calculated explicitly, however it 
is {\it different} from the almost-everywhere values (in the Lebesgue sense).
It can be calculated 
from $\lambda^G(x_{n_1})=2ln(1/x_{n_1})$. For the golden-mean number 
$x=(\sqrt{5}-1)/2$, for example, we have $n_1=1$ and 
$\lambda^G(x_{1})=2ln(1/x_{n_1})=0.96\ldots$ which is the lowest LE. All initial
conditions in $(0,1)$ which have the property 
$x=[n_1,n_2,n_3,\ldots,n_k,1,1,1,,\ldots]$, have the LE equal to $\sim 0.96$.
Period-2 orbits have the form 
$x=[n_1,n_2,n_1,n_2\ldots]= [\overline{n_1,n_2}]$, period-3 orbits
have $x=[\overline{n_1,n_2,n_3}]$, and so on. Note that for any rational $x$ 
the limit $j\rightarrow\infty$ from Eq.(\ref{LE}) does not exist and the LE 
cannot be determined.

 \begin{figure}[htb]
 \unitlength 1mm
 \begin{center}
 \includegraphics*[width=13.0cm,angle=0]{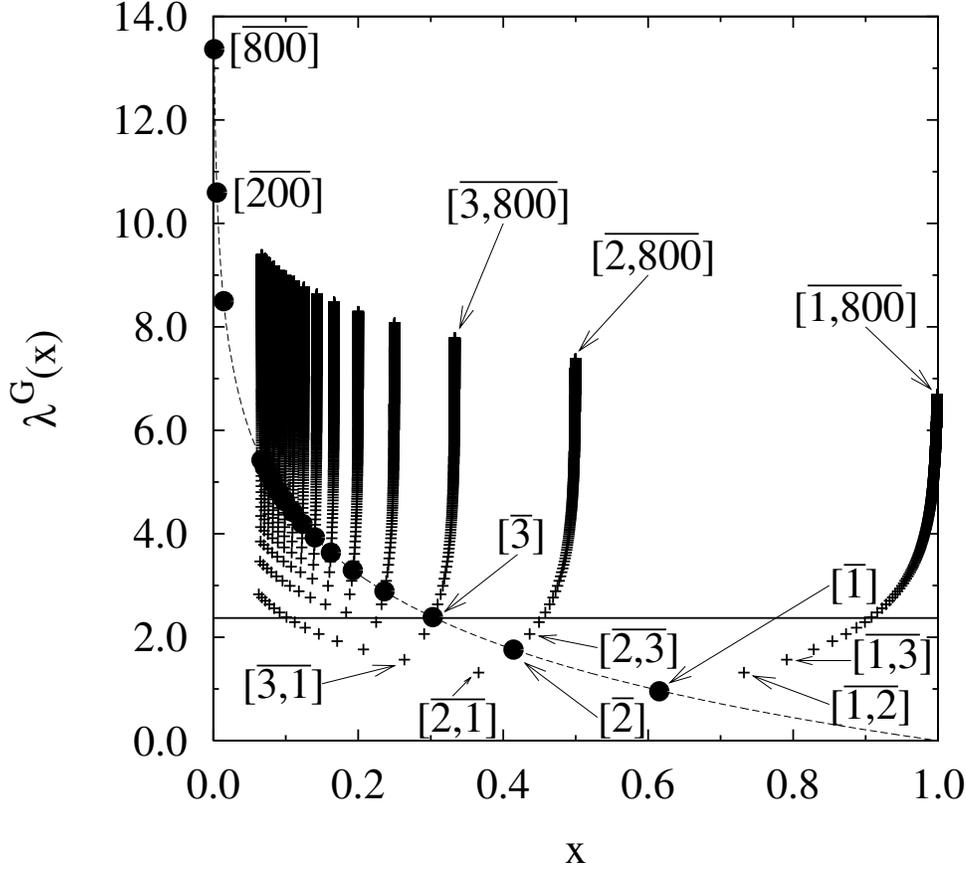}
 \end{center}
 \caption{Lyapunov exponents for period-1 and period-2 orbits 
from the Gauss map in the interval $(0,1)$. }
  \label{LEGauss}
  \end{figure}
Now we give a numerical summary of the above results for the LE of period-1 
and period-2 orbits in the Gauss map. If we iterate the Gauss map many times
($j\rightarrow\infty$)  
using arbitrary initial conditions between the interval $(0,1)$, all orbits 
have LEs equal to $\sim2.3731$, which is calculated using Eq.~(\ref{LE}). 
This is the ergodic result and it is shown by the solid line in 
Fig.~\ref{LEGauss}. Note that in this numerical procedure, due to numerical 
errors it is impossible to get exactly the POGM when $j\rightarrow\infty$.
As a consequence the LE is always $\sim2.3731$. However, for {\it low} 
period-p orbits we can calculate the LE which {\it differs} from this
ergodic results, since just few iterations are needed and numerical errors 
do not have time to propagate. We use the following procedure, we chose a 
period-p POGM map, determine the corresponding initial value of $x$ and then 
calculate numerically the corresponding LE $\lambda^G(x)$ using $j=p$. For 
period-1 orbits we have for $n_1=1,2,3,4$:

\begin{eqnarray}
x_1=0.618033989\ldots=[1,1,1,\ldots]=[\overline 1],\qquad 
\lambda^G(x_1)\sim0.96,\cr
& & \cr
x_2=0.414213562\ldots=[2,2,2,\ldots]=[\overline 2],
\qquad \lambda^G(x_2)\sim1.76,\cr
& & \cr
x_3=0.302775638\ldots=[3,3,3,\ldots]=[\overline 3],
\qquad \lambda^G(x_3)\sim2.38,\cr
& & \cr
x_4=0.236067977\ldots=[4,4,4,\ldots]=[\overline 4],
\qquad \lambda^G(x_4)\sim2.88.\cr
\nonumber
\label{per1}
\end{eqnarray}
We observe that using these value of $x_k$ ($k=1,2,3,4$), after one iteration 
of the Gauss map the LEs differ from the ergodic result. The LEs 
$\lambda^G(x_k)$
for period-1 points are plotted as circles in Fig.~\ref{LEGauss}. 
They are exactly on the dashed line, which is the curve $\ln{G^{\prime}(x)}$ 
for just one iteration. Points along the dashed line are the LE $\lambda^G(x)$
from the Gauss map  only when $x=x_1,x_2,\ldots$. Later we will 
explain the reason why we plotted the dashed line.

We can also calculate the LE for period-2 orbits. Some explicit example are
 
\begin{eqnarray}
x_{2,1}=0.366025\ldots\sim[\overline{2,1}],\qquad \lambda^G(x)\sim1.32,\cr
& & \cr
x_{3,1}=0.263762\ldots\sim[\overline{3,1}],\qquad \lambda^G(x)\sim1.56,\cr
& & \cr
x_{4,1}=0.207106\ldots\sim[\overline{4,1}],\qquad \lambda^G(x)\sim1.76.\cr
\nonumber
\label{per2}
\end{eqnarray} 
We generated the values of $x=[\overline{n_1,n_2}]$ for Period-2 orbits 
using all combinations of $n_1,n_2$. These points are marked as crossed 
points in Fig.~\ref{LEGauss}, where some corresponding CF representations 
are shown. Different sequences 
(branches) of $x$ values are observed. Sequences start (from below)
at $[\overline 1],[\overline{2,1}]$, $[\overline{3,1}],\ldots
[\overline{k,1}]\ldots,[\overline{800,1}]$. We could increase the last value 
of $k$ but it is not relevant for the discussion here.  In all simulations the 
last $k$ used was $k=800$. In Fig.~\ref{LEGauss} we showed only the $15$ first 
sequences, they come closer and closer as $k$ increases. Inside each sequence 
the CF representation changes its second number. For example, the first 
sequence, on the right, starts at the Golden mean $[\overline{1}]$ with 
the lowest LE for this sequence and the LE increases with period-2 orbits 
$[\overline{1,2}]$, $[\overline{1,3}]$ until $[\overline{1,800}]$, which 
has a LE $\sim6.69$. The next sequence
of period-2 orbits is $[\overline{2,1}]$, $[\overline{2,2}]$, 
$[\overline{2,3}],\ldots,[\overline{2,800}]$. The last point from this
sequence has a LE equal $\sim 7.37$. In fact, for 
$x_{2,k=\infty}=[\overline{2,\infty}]$, which is exactly equal $1/2$, the 
limit of (\ref{LE}) does not exist and the LE cannot be calculated.
Using the periodic orbit $x_{2,800}=[\overline{2,800}]\sim1/2$ we 
are allowed to calculate the LE $\lambda^G(x_{2,800})\sim 7.37$ very close (on 
the left ) to the non-periodic orbit $x=1/2$. 

The next sequence (see Fig.~\ref{LEGauss}) starts at 
$[\overline{3,1}]$ and ends at $x=[\overline{3,800}]\sim1/3$ with LE 
$\sim 7.78$. The subsequent sequences converge to $[\overline{4,800}]\sim1/4$
with $\lambda^G(x_{4,800})\sim 8.07$, $[\overline{5,800}]\sim1/5$
with $\lambda^G(x_{5,800})\sim 8.30\ldots$, $[\overline{800,800}]\sim 0.0012$
with $\lambda^G(x_{800,800})\sim 13.4$. It is important to observe that
the LE increases more and more as $k$ increases. For all these points the LE 
calculated for
the POGM {\it differ} from the ergodic result $\sim2.3731$. While such 
POGM have LM zero and may not be relevant for the Gauss map 
itself, we will show they contribution in a physical problem. We just need to 
relate the Gauss map and the 1D-Box HPC from section \ref{1Dpoint} through 
$x=\theta/\pi$.

\section{Yukawa interaction, results and discussion}
\label{1Dyukawa}

In order to study the stability of the ISs from the HPC
case, we need to apply a perturbation on the system. 
Therefore, we assume now that the interaction between particles is given by 
the Yukawa potential 

\begin{equation}
V(r)=V_0 {e^{-\alpha r}\over r},
\label{V}
\end{equation}
which has strength $V_0$ and the parameter $\alpha\ge0$ gives the interaction 
range $r_0=1/\alpha$. The classical dynamics of this 
problem was already analyse for equal masses \cite{ulloa} and for 
mass ratios $\gamma$ in the interval ($1,4$) \cite{nos}. 

\begin{figure}[htb]
 \unitlength 1mm
 \begin{center}
 \includegraphics*[width=13cm,angle=0]{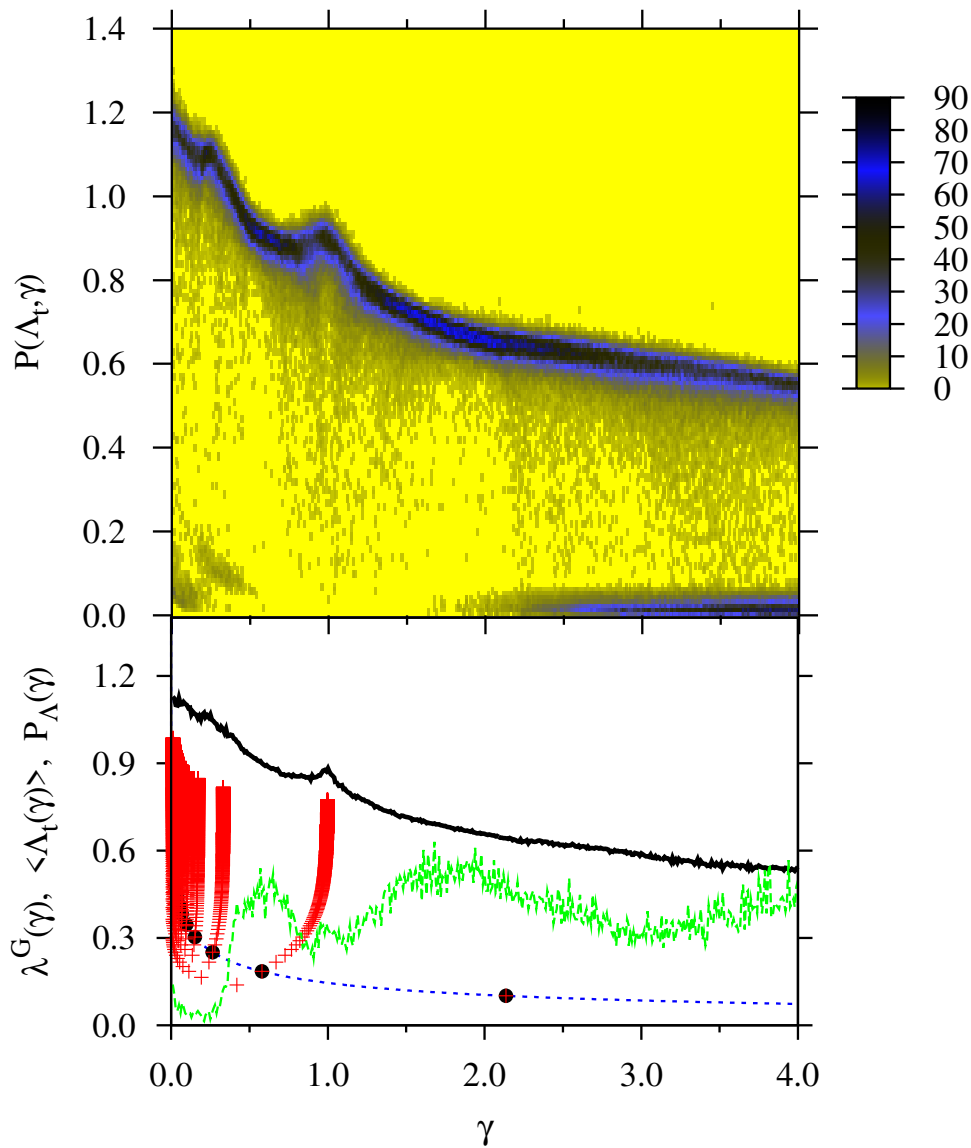}
 \end{center}
 \caption{a)Finite-time distribution of the largest Lyapunov exponent
 $P(\Lambda_{t},\gamma)$ calculated over $400$ trajectories up to time
 $t = 10^4$ and for $r_0\rightarrow\infty$.  With increasing
 $P(\Lambda_{t},\gamma)$ the color changes from light to dark (white
 over yellow and blue to black) and b) mean Lyapunov exponent (solid black 
 line), normalized distribution $P_{\Lambda}(\gamma)$ of the most
 probable Lyapunov exponent $\Lambda_{t}^{p}$ (solid gray line)
 and results for the LE from the Gauss map from Fig.~\ref{LEGauss}.}
  \label{mean}
  \end{figure}

Using the above interaction, we calculated the finite-time maximal LE as a 
function of the mass ratio in the interval $(0,4)$. Results are shown in 
Fig.~\ref{mean}(a) for
the  distribution $P(\Lambda_{t},\gamma)$ of the finite-time largest Lyapunov
exponents \cite{finite} $\Lambda_{t}$ and Fig.~\ref{mean}(b) for the mean 
$\langle\Lambda_{t}\rangle$ (see solid line). The mean LE decreases from 
roughly $1.18$ for $\gamma\sim0.0$ to $0.59$ for $\gamma=4.0$. An unexpected 
pronounced peak is observed at $\gamma\sim1.0$. When the value of $\gamma$ 
decreases, the mean LE exponent presents a minimum at $\gamma\sim 0.85$ and 
then increases again until another unexpected lower peak close to 
$\gamma\sim 0.11$ [better seen in Fig.~\ref{mean}a)]. 
The question now is: what is the origin of such peaks? 
Why are there special values of $\gamma$ where the motion is more chaotic?

Figure \ref{mean}(b) also shows the results for the $\gamma$ dependence of the
LE from the Gauss map showed in Fig.~\ref{LEGauss}.  We used relations 
$\theta=\pi x$ and $\gamma(x)={\frac{1-\cos{\pi x}}{1+\cos{\pi x}}}$ from
Eq.~(\ref{cost}) to calculate the correspondence
between $\gamma$ and $x$ from the Gauss map. Note that in this interval
of $\gamma$ only the Golden mean point $[\overline 1]$ remains from the 
first sequence (on the right) from Fig.~\ref{LEGauss}. All other points
from this sequence have $\gamma>4.0$. First observation is that the main 
qualitative behavior of the LE from our physical model follows the curve 
calculated for the first iteration from the Gauss map (see dashed line).
In other words, the main qualitative behavior of the mean LE from the 
1D-box with YI follows the period-1 POGM. 
Second observation is that the pronounced peak observed at $\gamma\sim1.0$ is 
{\it very close} to the point $\gamma(x=[\overline{2,800}]\sim1/2)$
which is a period-2 orbit 
from the Gauss map. We have to remember that this point is very close to 
the non-periodic orbit $x=1/2$ from the Gauss map where the LE cannot be 
calculated. The other pronounced (lower) peak at $\gamma\sim 0.17$ is close 
to the period-2 orbit $\gamma(x=[\overline{4,800}]\sim1/4)$ from the Gauss 
map. As the value of $\gamma$ decreases, the LE from the 1D-box with YI follows
all peaks obtained from the POGM $x(\gamma)=[\overline{k,800}]$ 
($k=2,3,\ldots,800$). 
The important point is, the LE from our physical model follows
the behavior of the LE from the LM zero POGM. Therefore, both 
pronounced peaks at $\gamma\sim1.0, 0.17$ are probably related to 
signatures from the {\it periodic orbits of the HPC}, for 
which the LE cannot be estimated using the Gauss map.

Another interesting feature appears if we calculate the change of the 
width of $P(\Lambda_{t},\gamma)$ around the most probable $\Lambda^{p}_{t}$ 
defined through

\begin{equation}
    \label{probable}
   \left.\frac{\partial
P(\Lambda_{t},\gamma)}{\partial
\Lambda_{t}}\right|_{\Lambda_{t}=\Lambda_{t}^{p}}\,=0.
\end{equation}
This quantity, called $P_{\Lambda}(\gamma)$, has been proposed \cite{nos} 
as a sensitive measure of `stickness' in phase-space which are a consequence 
of the existence of regular islands. Each time this quantity has a minimum, 
the mixed phase-space of a system is expected to have more trapped 
trajectories. Otherwise it has an ``ergodic-like'' motion, 
then mostly all initial conditions converge to the same finite time LE. For 
more details about this quantity, some examples and motivations, we refer the 
readers to \cite{nos}. This quantity is plotted in Fig.~\ref{mean}(b) (see 
gray curve with strong variations). Clearly three minima are observed 
close to $\gamma\sim0.25,1.0,3.0$ where the dynamics in phase-space has more 
trapped (`sticky') trajectories. As shown in another work \cite{nos}, the 
values $\gamma\sim1.0,3.0$ can be related to the integrable cases (genus 
$g=1$, 
see Table \ref{tab}) from the HPC case. Here we show additional results for 
mass ratio in the interval ($0.0,1.0$). Besides for $\gamma=3.0$, it is very 
interesting to observe that $P_{\Lambda}(\gamma)$ has a minimum for all
points for which $\gamma(x=[\overline{k,800}])$ ($k=2,3,\ldots,800$) from 
the Gauss map. As a consequence,  $P_{\Lambda}(\gamma)$ has a minimum in the 
extended interval $\gamma\sim(0.0,0.4)$, where these points come closer and 
closer. Since the points $\gamma(x=[\overline{k,800}])$  ($k=2,3,\ldots,800$) 
are very close to the non-POGM (which are the PO from the model), it suggest 
that the 
minimum of  $P_{\Lambda}(\gamma)$ are due the trapped trajectories reminiscent 
from the PO from the physical model (with HPC or YI).

Above results allow us to make some statements about the stability of the 
linear unstable dynamics from the HPC case, i.~e., the stability from the
different topological surfaces present in the HPC case. Table \ref{tab} gives 
some examples of the relation between the values of $x=\theta/\pi$ with the LE 
$\lambda^G(x_{k,800})$ from the Gauss map calculated {\it very} close (on the
left) to these points. For example,  $\theta/\pi=1/2$($\gamma=1$) is an 
integrable case from the HPC problem with genus $g=1$ (torus). The LE for the 
POGM calculated at $\gamma(x=[\overline{2,800}])$ gives 
$\lambda^G(x_{2,800})\sim7.37$. We are proposing that this value of the LE 
gives a possible ``degree of instability'' of the torus with genus $g=1$. We 
observe in Table \ref{tab} that as we increase $g$, the
corresponding values of $\lambda^G(x_{k,800})$ also increase and $\gamma$
decreases. Therefore, we expect that for higher values of $g$ the invariant
surfaces from the HPC case are more unstable. This is verified for the problem 
considered in this paper (see Fig.~\ref{mean}), where the mean LE increases
as the genus $g$ increases following values from Table \ref{tab}.

\section{Conclusions}
\label{conclusion}

While the KAM theorem \cite{KAM} makes some statements about the existence 
of non-periodic orbits (KAM curves), the residue criterion 
\cite{Greene2} establishes a correspondence between the existence of a KAM 
curve and the stability of the periodic orbits that approximate it. Such
criterion can be used to determine for which parameter of the model the KAM
curves may be destroyed. Here the 
stability of {\it periodic orbits with LM zero} from the Gauss map (i.~e., the 
{\it non-periodic curves} from the physical model) allows us to get some 
insight about the stability of ISs (no KAM curves) from a linear unstable 
system with LE {\it zero} . The linear unstable system considered here is the 
1D-box 
with two particles interacting via HPC. The stability of the IS  is obtained by
calculating the LE (via Gauss map) from the CF representation of the masses 
ratio. By perturbing the IS with a Yukawa interaction between particles, 
we observe that the LEs follow qualitatively the LEs from the Gauss map
[see Fig.~\ref{mean}b)]. Only periodic orbits with LM zero from the Gauss map 
seems to be relevant. Additionally, the two more pronounced peaks in the mean 
LE [see Fig.~\ref{mean}a)-b)] at $\gamma\sim1.0$ and $\sim0.17$
are explained with results from the Gauss map.
They are probably related to signatures from the {\it periodic orbits of the 
HPC} case, for which the LE cannot be estimated via Gauss map. This shows a 
nice example of the relation between the ``instability'' of a simple CF 
representation of a number to the LEs from a physical model. We also show
that pseudo-integrable systems, where the IS has a higher genus $g$, 
are more unstable under perturbations. 
This is easy to see from Table \ref{tab} and Fig.~\ref{mean}b). As the value 
of $g$ increases, the corresponding value of $\gamma$ decreases and the LE 
increases. Moreless, we were able to show that `stickness' effects are 
present each 
time the mass ratio $\gamma$ is close to POGM and therefore, close to the 
non-periodic orbits from the 1D box with HPC. This was quantified by 
calculating $P_{\Lambda}(\gamma)$, which is a sensitive measure of trapped 
trajectories in phase-space. Each time $P_{\Lambda}(\gamma)$ has a minimum
[see gray line in Fig.~\ref{mean}b)], trapped trajectories in phase-space are 
expected .

\vspace*{1cm}
\section*{Acknowledgments}
\noindent
CM and MWB thank CNPq for financial support. 
MWB is greatfull to J.~M.~Rost for helpful discussions.

\end{document}